\documentclass[runningheads]{llncs}

\usepackage{hyperref}       
\usepackage{url}            
\usepackage{booktabs}       
\usepackage{amsfonts}       
\usepackage{nicefrac}
\usepackage{lipsum}
\usepackage{listings}
\usepackage{graphicx}
\usepackage{epstopdf}
\usepackage{array}
\usepackage{subcaption}

\usepackage{amsfonts}
\usepackage{mathtools}
\usepackage{cite}

\usepackage{stackengine,graphicx}
\stackMath

\usepackage{amssymb}
\usepackage{marvosym,enumerate,color,mathrsfs,graphicx,epstopdf}
\usepackage{enumitem}
\setenumerate{listparindent=\parindent}

\usepackage[dvipsnames]{xcolor}
\usepackage[normalem]{ulem}
\usepackage{bm}
\usepackage{mathtools}
\usepackage{mathrsfs}
\usepackage{verbatim}
\usepackage[utf8]{inputenc}
\usepackage{algorithm}
\usepackage{algorithmic}

\newcommand{\xbf}{\mathbf{x}}

\newcommand{\rbf}{\mathbf{r}}
\newcommand{\bbf}{\mathbf{b}}
\newcommand{\ubf}{\mathbf{u}}
\newcommand{\vbf}{\mathbf{v}}
\newcommand{\fbf}{\mathbf{f}}
\newcommand{\sbf}{\mathbf{s}}
\newcommand{\Fbf}{\mathbf{F}}
\newcommand{\Pbf}{\mathbf{P}}

\DeclareMathOperator{\prox}{\mathrm{prox}}

\newcommand{\R}{\mathbb{R}}

\DeclareMathOperator*{\argmin}{arg\,min}

\newcommand{\J}{\mathcal{J}}

\newcommand{\soft}{\mathcal{S}}

\newcommand{\G}{\mathcal{G}}

\begin{document}
\title{Deep Parallel MRI Reconstruction Network Without Coil Sensitivities}

\author{
Wanyu Bian\inst{1} \and
Yunmei Chen\inst{1} \and
Xiaojing Ye\inst{2}}

\authorrunning{W. Bian et al.}
%
\institute{University of Florida, Gainesville FL 32611, USA\\
\email{ \{wanyu.bian, yun\}@ufl.edu} \and
Georgia State University, Atlanta GA 30302, USA\\
\email{xye@gsu.edu}}

\maketitle

\begin{abstract}
We propose a novel deep neural network architecture by mapping the robust proximal gradient scheme for fast image reconstruction in parallel MRI (pMRI) with regularization function trained from data. The proposed network learns to adaptively combine the multi-coil images from incomplete pMRI data into a single image with homogeneous contrast, which is then passed to a nonlinear encoder to efficiently extract sparse features of the image. Unlike most of existing deep image reconstruction networks, our network does not require knowledge of sensitivity maps, which can be difficult to estimate accurately, and have been a major bottleneck of image reconstruction in real-world pMRI applications. The experimental results demonstrate the promising performance of our method on a variety of pMRI imaging data sets.
\end{abstract}

\section{Introduction}
\label{sec:intro}
Parallel magnetic resonance imaging (pMRI) is a state-of-the-art medical MR imaging technology which surround the scanned objects by multiple receiver coils and collect k-space (Fourier) data in parallel. To accelerate scan process, partial data acquisitions that increase the spacing between read-out lines in k-space are implemented in pMRI. However, reduction in k-space data sampling arising aliasing artifacts in images, which must be removed by image reconstruction process.
There are two major approaches to image reconstruction in pMRI: the first approach are k-space methods which interpolate the non-sampled k-space data using the sampled ones across multiple receiver coils \cite{doi:10.1002/jmri.23639}, such as the generalized auto-calibrating partially parallel acquisition (GRAPPA) \cite{griswold2002generalized}. The other approach is the class of image space methods which remove the aliasing artifacts in the image domain by solving a system of equations that relate the image to be reconstructed and partial k-spaced data through coil sensitivities, such as in SENSitivity Encoding (SENSE) \cite{pruessmann1999sense}.

In this paper, we propose a new deep learning based reconstruction method to address several critical issues of pMRI reconstruction in image space. Consider a pMRI system with $N_c$ receiver coils acquiring 2D MR images at resolution $m\times n$ (we treat a 2D image $\vbf \in \mathbb{C}^{m\times n}$ and its column vector form $\vbf \in \mathbb{C}^{mn}$ interchangeably hereafter). Let $\Pbf \in \mathbb{R}^{p\times mn}$ be the binary matrix representing the undersampling mask with $p$ sample locations in k-space, and $\sbf_i\in\mathbb{C}^{mn}$ the coil sensitivity and $\fbf_i \in \mathbb{C}^{p}$ the \emph{partial} k-space data at the $i$th receiver coil for $i=1,\dots,N_c$. Therefore $\fbf_i$ and the image $\vbf$ are related by $\fbf_i = \Pbf \Fbf (\sbf_i \cdot \vbf) + \mathbf{n}_i$ where $\cdot$ denotes pointwise multiplication of two matrices, and $\mathbf{n}_i$ is the unknown acquisition noise in k-space at each receiver coil. Then SENSE-based image space reconstruction methods can be generally formulated as an optimization problem:
\begin{equation}\label{eq:PFS}
    \min_{\vbf} \ \sum^{N_c}_{i=1} \frac{1}{2} \| \Pbf \Fbf (\sbf_i \cdot \vbf)- \fbf_i\|^2 + R(\vbf),
\end{equation}
where $\vbf\in \mathbb{C}^{m n}$ is the MR image to be reconstructed, $\Fbf \in \mathbb{C}^{mn\times mn}$ stands for the discrete Fourier transform, and $R(\vbf)$ is the regularization on the image $\vbf$.
 $\| \xbf \|^2 := \|\xbf \|_2^2 = \sum_{j=1}^n |x_j|^2$ for any complex vector $\xbf = (x_1,\dots,x_n)^{\top} \in \mathbb{C}^n$.
There are two critical issues in pMRI image reconstruction using \eqref{eq:PFS}: availability of accurate coil sensitivities $\{\sbf_i\}$ and proper image regularization $R$.
Most existing SENSE-based reconstruction methods assume coil sensitivity maps are given, which are however difficult to estimate accurately in real-world applications. 
On the other hand, the regularization $R$ is of paramount importance to the inverse problem \eqref{eq:PFS} to produce desired images from significantly undersampled data, but a large number of existing methods employ handcrafted regularization which are incapable to extract complex features from images effectively.

In this paper, we tackle the two aforementioned issues in an unified deep-learning framework dubbed as pMRI-Net.
Specifically, we consider the reconstruction of multi-coil images $\ubf = (\ubf_1,\dots,\ubf_{N_c}) \in \mathbb{C}^{mnN_c}$ for all receiver coils to avoid use of coil sensitivity maps (but can recover them as a byproduct), and design a deep residual network  which can jointly learn the adaptive combination of multi-coil images and an effective regularization from training data.

The contribution of this paper could be summarized as follows:
Our method is the first ``combine-then-regularize" approach for deep-learning based pMRI image reconstruction. The combination operator integrates multichannel images into single channel and this approach performs better than the linear combination the root of sum-of-squares (SOS) method \cite{pruessmann1999sense}.
This approach has three main advantages: (i) the combined image has homogeneous contrast across the FOV, which makes it suitable for feature-based image regularization and less affected by the intensity biases in coil images; (ii) the regularization operators are applied to this single body image in each iteration, and require much fewer network parameters to reduce overfitting and improve robustness; and (iii) our approach naturally avoids the use of sensitivity maps, which has been a thorny issue in image-based pMRI reconstruction.

\section{Related Work}
\label{sec:related}
Most existing deep-learning (DL) based methods rendering end-to-end neural networks mapping from the partial k-space data to the reconstructed images \cite{WANG2020136,7493320,doi:10.1002/mp.12600,Quan2018CompressedSM,8417964}. The common issue with this class of methods is that the DNNs require excessive amount of data to train, and the resulting networks perform similar to ``black-boxes'' which are difficult to interpret and modify.

In recent years, a class of DL based methods improve over the end-to-end training by selecting the scheme of an iterative optimization algorithm and prescribe a phase number $T$, map each iteration of the scheme to one phase of the network. These methods are often known as the learned optimization algorithms (LOAs),  \cite{Aggarwal_2019,10.1007/978-3-030-32248-9_3,doi:10.1002/mrm.26977,8550778,NIPS2016_6406,zhang2018ista,8067520}.
For instance, ADMM-Net \cite{NIPS2016_6406},  ISTA-Net$^+$ \cite{zhang2018ista}, and cascade network \cite{8067520} are regular MRI reconstruction.
For pMRI: Variational network (VN)\cite{doi:10.1002/mrm.26977} introduced gradient descent method by applying given sensitivities $\{\sbf_i\}$. MoDL \cite{Aggarwal_2019} proposed a recursive network by unrolling the conjugate gradient algorithm using a weight sharing strategy.
 Blind-PMRI-Net \cite{10.1007/978-3-030-32251-9_80}  designed three network blocks to alternately update multi-channel images, sensitivity maps and the reconstructed MR image using an iterative algorithm based on half-quadratic splitting. The network in \cite{10.1007/978-3-030-32248-9_5} developed a Bayesian framework for joint MRI-PET reconstruction. VS-Net \cite{10.1007/978-3-030-32251-9_78} derived a variable splitting optimization method. However, existing methods still face the lack of accurate coil sensitivity maps and proper regularization in the pMRI problem.

Recently, a method called  DeepcomplexMRI \cite{WANG2020136} developed an end-to-end learning without explicitly using coil sensitivity maps to recover channel-wise images, and then combine to a single channel image in testing.

This paper proposes a novel deep neural network architecture which integrating the robust proximal gradient scheme for pMRI reconstruction without knowledge of coil sensitivity maps. Our network learns to adaptively combine the channel-wise image from the incomplete data to assist the reconstruction and learn a nonlinear mapping to efficiently extract sparse features of the image by using a set of training data on the pairs of under-sampled channel-wise k-space data and corresponding images. The roles of the multi-coil image combination operator and sparse feature encoder are clearly defined and jointly learned in each iteration. As a result, our network is more data efficient in training and the reconstruction results are more accurate.

\section{Proposed Method}
\label{sec:proposed}
\subsection{Joint image reconstruction pMRI without coil sensitivities}
We propose an alternative pMRI reconstruction approach to \eqref{eq:PFS} by recovering images from individual receiver coils jointly.
Denote $\ubf_i$ the MR image at the $i$th receiver coil, i.e., $\ubf_i = \sbf_i \cdot \vbf$, where the sensitivity $\sbf_i$ and the full FOV image $\vbf$ are both unknown in practice.
Thus, the image $\ubf_i$ relates to the partial k-space data $\fbf_i$ by $\fbf_i = \Pbf \Fbf \ubf_i + \mathbf{n}_i$, and hence the data fidelity term is formulated as least squares $(1/2) \cdot \|\Pbf \Fbf \ubf_i - \fbf_i\|^2$.
We also need a suitable regularization $R$ on the images $\{\ubf_i\}$.
However, these images have vastly different contrasts due to the significant variations in the sensitivity maps at different receiver coils.
Therefore, it is more appropriate to apply regularization to the (unknown) image $\vbf$.

To address the issue of regularization, we propose to first learn a nonlinear operator $\J$ that combines $\{\ubf_i\}$ into the image $\vbf = \J (\ubf_1,\dots,\ubf_{N_c}) \in \mathbb{C}^{m\times n}$ with homogeneous contrast, and apply a regularization on $\vbf$ with a parametric form $\| \G (\vbf) \|_{2,1}$ by leveraging the robust sparse selection property of $\ell_{2,1}$-norm and shrinkage threshold operator. Here $\G(\vbf)$ represents a nonlinear sparse encoder trained from data to effectively extract complex features from the image $\vbf$.
Combined with the data fidelity term above, we propose the following pMRI image reconstruction model:
\begin{equation}\label{eq:m}
\ubf(\fbf; \Theta) = \argmin_{\ubf}\ \frac{1}{2} \sum^{N_c}_{i=1} \| \textbf{PF} \ubf_i - \fbf_i \|^2_2  +  \| \G \circ \J  (\ubf)\|_{2,1},
\end{equation}
where $\ubf = (\ubf_1,\dots,\ubf_{N_c})$ is the multi-channel image to be reconstructed from the pMRI data $\fbf=(\fbf_1,\dots,\fbf_{N_c})$, and $\Theta=(\G,\J)$ represents the parameters of the deep networks $\G$ and $\J$.
The key ingredients of \eqref{eq:m} are the nonlinear combination operator $\J$ and sparse feature encoder $\G$, which we describe in details in Section \ref{subsec:network}.
Given a training data set consisting of  $J$ pairs $\{ (\fbf^{[j]}, \hat{\ubf}^{[j]} ) \,|\, \ 1\le j\le J \}$, where $\fbf^{[j]}=(\fbf^{[j]}_1,\dots,\fbf^{[j]}_{N_c})$ and $\hat{\ubf}^{[j]}=(\hat{\ubf}^{[j]}_1,\dots,\hat{\ubf}^{[j]}_{N_c})$ are respectively the partial k-space data and the ground truth image reconstructed by full k-space data of the $j$th image data, our goal is to learn $\Theta$ (i.e., $\G$ and $\J$) from the following bi-level optimization problem:
\begin{equation}
    \label{eq:bilevel}
    \min_{\Theta} \frac{1}{J} \sum_{j=1}^J \ell(\ubf(\fbf^{[j]};\Theta), \hat{\ubf}^{[j]}),\ \mbox{s.t.}\ \ubf(\fbf^{[j]};\Theta)\ \mbox{solves \eqref{eq:m} with data $\fbf^{[j]}$},
\end{equation}
where $\ell(\ubf,\hat{\ubf})$ measures the discrepancy between the reconstruction $\ubf$ and the ground truth $\hat{\ubf}$.
To tackle the lower-level minimization problem in \eqref{eq:bilevel}, we construct a proximal gradient network with residual learning as an (approximate) solver of \eqref{eq:m}.
Details on the derivation of this network are provided in the next subsection.

\subsection{Proximal gradient network with residual learning}
\label{subsec:pg}
If the operators $\J$ and $\G$ were given, we can apply proximal gradient descent algorithm to approximate a (local) minimizer of \eqref{eq:m} by iterating
\begin{subequations}\label{eq:bui}
\begin{align}
\bbf_i^{(t)} & =  \ubf_i^{(t)} - \rho_t \Fbf^{\top}  \Pbf^{\top} (\Pbf \Fbf \ubf_i^{(t)} - \fbf_i), \label{eq:bi}  \\
\ubf_i^{(t+1)} &  = [\prox_{\rho_t\|\G\circ \J (\cdot) \|_{2,1}}(\bbf^{(t)})]_i, \quad 1\le i \le N_c \label{eq:ui} \end{align}
\end{subequations}
where $\bbf^{(t)} = (\bbf_1^{(t)},\dots,\bbf_{N_c}^{(t)})$, $[\xbf]_i = \xbf_i \in \mathbb{C}^{mn}$ for any vector $\xbf \in \mathbb{C}^{mn N_c}$, $\rho_t>0$ is the step size, and $\prox_{g}$ is the proximal operator of $g$ defined by
\begin{equation}
    \prox_{g}(\bbf) = \argmin_{\xbf} g(\xbf) + \frac{1}{2} \| \xbf - \bbf \|^2.
\end{equation}
The gradient update step \eqref{eq:bi} is straightforward to compute and fully utilizes the relation between the partial k-space data $\fbf_i$ and the image $\ubf_i$ to be reconstructed as derived from MRI physics.
The proximal update step \eqref{eq:ui}, however, presents several difficulties: the operators $\J$ and $\G$ are unknown and need to be learned from data, and the proximal operator $\prox_{\rho_t\|\G\circ \J (\cdot) \|_{2,1}}$ most likely will not have closed form and can be difficult to compute.
Assuming that we have both $\J$ and $\G$ parametrized by convolutional networks, we adopt a residual learning technique by leveraging the shrinkage operator (as the proximal operator of $\ell_{2,1}$-norm $\|\cdot \|_{2,1}$) and converting \eqref{eq:ui} into an explicit update formula.
To this end, we parametrize the proximal step \eqref{eq:ui} as an implicit residual update:
\begin{equation}\label{eq:ui_update}
    \ubf_i^{(t+1)} = \bbf_i^{(t)} + [\rbf(\ubf_1^{(t+1)}, \cdots, \ubf_{N_c}^{(t+1)})]_i,
\end{equation}
where $\rbf=\tilde{\J} \circ \tilde{\G} \circ \G \circ \J$ is the residual network as the composition of $\J$, $\G$, and their adjoint operators $\tilde{\J}$ and $\tilde{\G}$. These four operators are learned separately to increase the capacity of the network.
To reveal the role of nonlinear shrinkage selection in \eqref{eq:ui_update}, consider the original proximal update \eqref{eq:ui} where
\begin{equation}\label{eq:u_prox}
    \ubf^{(t+1)} = \argmin_{\ubf} \|\G \circ \J (\ubf) \|_{2,1} + \frac{1}{2 \rho_t} \| \ubf - \bbf^{(t)}\|^2.
\end{equation}
For certain convolutional networks $\J$ and $\G$ with rectified linear unit (ReLU) activation, $\| \ubf - \bbf^{(t)}\|^2$ can be approximated by $\alpha \| \G\circ \J(\ubf) - \G\circ \J(\bbf^{(t)})\|^2$ for some $\alpha > 0$ dependent on $\J$ and $\G$ \cite{zhang2018ista}.
Substituting this approximation into \eqref{eq:u_prox}, we obtain that
\begin{equation}\label{eq:GJu}
    \G \circ \J (\ubf^{(t+1)}) = \soft_{\alpha_t} ( \G \circ \J (\bbf^{(t)})),
\end{equation}
where $\alpha_t = \rho_t/\alpha$, $\soft_{\alpha_k}(\xbf) = \prox_{\alpha_k \|\cdot \|_{2,1}} (\xbf) = [\mathrm{sign}(x_i) \max(|x_i| - \alpha_k, 0)] \in \mathbb{R}^n$ for any vector $\xbf = (x_1,\dots,x_n)\in \mathbb{R}^n$ is the soft shrinkage operator.
Plugging \eqref{eq:GJu} into \eqref{eq:ui_update}, we obtain an explicit form of \eqref{eq:ui}, which we summarize together with \eqref{eq:bi} in the following scheme:
\begin{subequations}\label{eq:bu}
\begin{align}
\bbf_i^{(t)} & =  \ubf_i^{(t)} - \rho_t \Fbf^{\top}  \Pbf ^{\top} (\Pbf \Fbf \ubf_i^{(t)} - \fbf_i), \label{eq:b}  \\
\ubf_i^{(t)} & = \bbf_i^{(t)} + [\tilde{\J} \circ \tilde{\G} \circ \soft_{\alpha_t} ( \G \circ \J (\bbf^{(t)}))]_i, \quad 1\le i \le N_c
\label{eq:u}
\end{align}
\end{subequations}
Our proposed reconstruction network thus is composed of a prescribed $T$ phases, where the $t$th phase performs the update of \eqref{eq:bu}.
With a zero initial  $ \{ \ubf_i^{(0)} \} $ and partial k-space data $\{\fbf_i\}$ as input, the network performs the update \eqref{eq:bu} for $1\le t\le T$ and finally outputs $\ubf^{(T)}$.
This network serves as a solver of \eqref{eq:m} and uses $\ubf^{(T)}$ as an approximation of the true solution $\ubf(\fbf;\Theta)$. Hence, the constraint in \eqref{eq:bilevel} is replaced by this network for every input data $\fbf^{[j]}$.

\subsection{Network architectures and training}\label{subsec:network}
We set $\J$ as a convolutional network with $N_l=4$ layers and each linear convolution of kernel size $3 \times 3$. The first $N_l-1$ layers have $N_f=64$ filter kernels, and $N_f=1$ in the last layer. Each layer follows an activation ReLU except for the last layer. The operator $\G$ being set as the same way except that $N_f=32$ and kernel size is $ 9\times 9$.
Operators $\tilde{ \J}$ and $\tilde{ \G }$ are designed in symmetric structures as $\J$ and $ \G$ respectively.
We treat a complex tensor as a real tensor of doubled size, and apply convolution separately.  More details on network structure are provided in Supplementary Material.

The training data $(\fbf,\hat{\ubf})$ consists of $J$ pairs $\{( \fbf_{i}^{[j]}, \hat{\ubf}_{i}  ^{[j]}) \,|\, 1 \le i \le N_c,\ 1\le j\le J \}$.   
To increase network capacity, we allow varying operators of \eqref{eq:bu} in different phases.
Hence  $\Theta = \{ \rho_t,\alpha_t,\J^{(t)},\G^{(t)},\tilde{ \G}^{(t)}, \tilde{ \J}^{(t)}\,|\, 1\le t \le T \}$ are the parameters to be trained.
Based on the analysis of loss functions \cite{7797130, cc2825e6939c4d09981c5d9c7ce29823, doi:10.1002/mp.13628}, the optimal parameter $ \Theta$ can be solved by minimizing the loss function:
We set the discrepancy measure $\ell$ between the reconstruction $\ubf$ and the corresponding ground truth $\hat{\ubf}$ in \eqref{eq:bilevel} as follows,
\begin{equation}
\label{eq:loss}
\begin{aligned}
\ell(\ubf,\hat{\ubf}) = \| \sbf(\ubf) - \sbf(\hat{\ubf})\|_2 + \gamma \| |\J(\ubf)| - \sbf(\hat{\ubf})\|_2
\end{aligned}
\end{equation}
where $\sbf(\ubf)= (\sum_{i=1}^{N_c} |\ubf_i|^2)^{1/2} \in\mathbb{R}^{mn}$ is the pointwise root of sum of squares across the $N_c$ channels of $\ubf$, $|\cdot|$ is the pointwise modulus, and $\gamma>0$ is a weight function. We also tried replacing the first by $(1/2)\cdot \|\ubf - \hat{\ubf}\|_2^2$, but it seems that the one given in \eqref{eq:loss} yields better results in our experiments. The second term of \eqref{eq:loss} can further improve accuracy of the magnitude of the reconstruction. The initial guess (also the input of the reconstruction network) of any given pMRI $\fbf^{[j]}$ is set to the zero-filled reconstruction $ \Fbf^{-1} \fbf^{[j]}$, and the multi-channel image $\ubf^{(T)}(\fbf^{[j]}; \Theta)$ is the output of the network \eqref{eq:bu} after $T$ phases. In addition, $  \J( {\ubf}^{(T)}(\fbf^{[j]}; \Theta))$ is the final single body image reconstructed as a by-product (complex-valued).

\section{Experimental Results}
\begin{table}[t]
\centering
\caption{Quantitative measurements for reconstruction of Coronal FSPD data. } \label{tab:fs}
\begin{tabular}{cccc}
\toprule
Method   & PSNR                    & SSIM              & RMSE          \\\midrule
GRAPPA\cite{griswold2002generalized}   & 24.9251$\pm$0.9341    & 0.4827$\pm$0.0344  & 0.2384$\pm$0.0175 \\
SPIRiT\cite{doi:10.1002/mrm.22428}  & 28.3525$\pm$1.3314    & 0.6509$\pm$0.0300  & 0.1614$\pm$0.0203 \\
VN\cite{doi:10.1002/mrm.26977}       & 30.2588$\pm$1.1790    & 0.7141$\pm$0.0483  & 0.1358$\pm$0.0152 \\
DeepcomplexMRI\cite{WANG2020136} & 36.6268$\pm$1.9662   & 0.9094$\pm$0.0331 & 0.0653$\pm$0.0085 \\
pMRI-Net ~&~ \textbf{37.8475$\pm$1.2086} ~&~ \textbf{0.9212$\pm$0.0236} ~&~ \textbf{0.0568$\pm$0.0069}~ \\ \bottomrule
\end{tabular}
\end{table}
\label{sec:experiment}
 \begin{table}[t]
\centering
\caption{Quantitative measurements for reconstruction of Coronal PD data. } \label{tab:pd}
\begin{tabular}{cccc}
\toprule
Method         & PSNR                & SSIM                & RMSE           \\ \midrule
GRAPPA\cite{griswold2002generalized} & 30.4154$\pm$0.5924  & 0.7489$\pm$0.0207   & 0.0984$\pm$0.0030 \\
SPIRiT\cite{doi:10.1002/mrm.22428}   & 32.0011$\pm$0.7920  & 0.7979$\pm$0.0306   & 0.0824$\pm$0.0082 \\
VN \cite{doi:10.1002/mrm.26977}              & 37.8265$\pm$0.4000    & 0.9281$\pm$0.0114    & 0.0422$\pm$0.0036 \\
DeepcomplexMRI \cite{WANG2020136} & 41.5756$\pm$0.6271  & 0.9679$\pm$0.0031   & 0.0274$\pm$0.0018 \\
pMRI-Net   ~&~ \textbf{42.4333$\pm$0.8785} ~&~ \textbf{0.9793$\pm$0.0023} ~&~ \textbf{0.0249$\pm$0.0024}\\ \bottomrule
\end{tabular}
\end{table}

\begin{figure*}
\centering
\includegraphics[width=0.16\linewidth, angle=180]{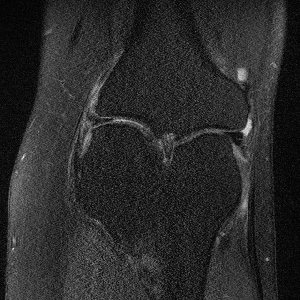}
\includegraphics[width=0.16\linewidth, angle=180]{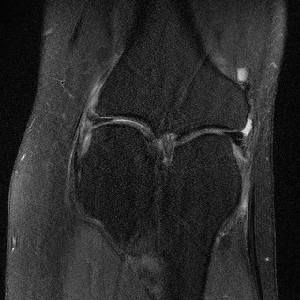}
\includegraphics[width=0.16\linewidth, angle=180]{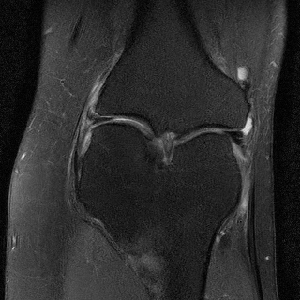}
\includegraphics[width=0.16\linewidth, angle=180]{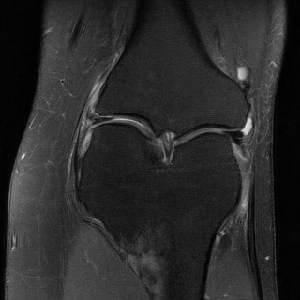}
\includegraphics[width=0.16\linewidth, angle=180]{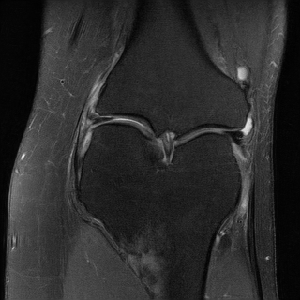}
\includegraphics[width=0.16\linewidth, angle=180]{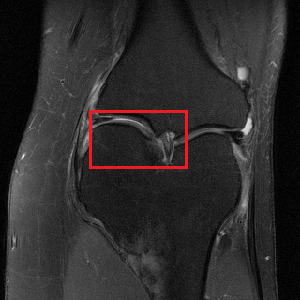}\\
\includegraphics[width=0.16\linewidth, angle=180]{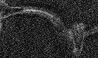}
\includegraphics[width=0.16\linewidth, angle=180]{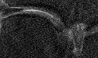}
\includegraphics[width=0.16\linewidth, angle=180]{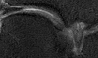}
\includegraphics[width=0.16\linewidth, angle=180]{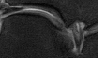}
\includegraphics[width=0.16\linewidth, angle=180]{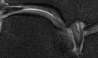}
\includegraphics[width=0.16\linewidth, angle=180]{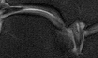}\\
\includegraphics[width=0.16\linewidth, angle=180]{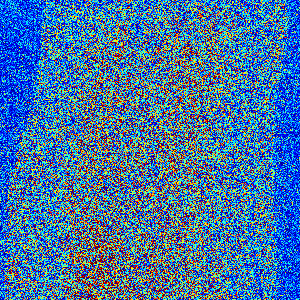}
\includegraphics[width=0.16\linewidth, angle=180]{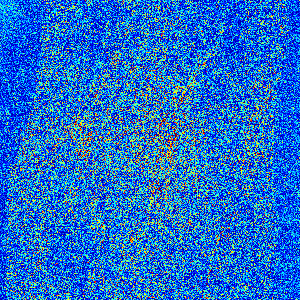}
\includegraphics[width=0.16\linewidth, angle=180]{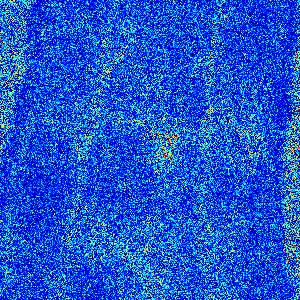}
\includegraphics[width=0.16\linewidth, angle=180]{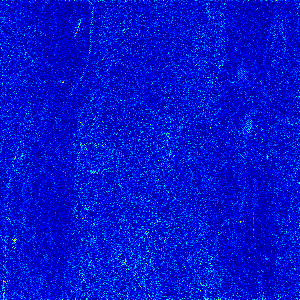}
\includegraphics[width=0.16\linewidth, angle=180]{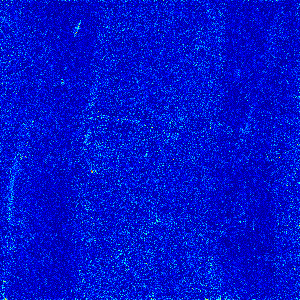}
\includegraphics[width=0.16\linewidth, angle=180]{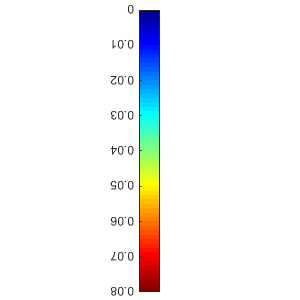}
\caption{Results on the Coronal FSPD knee image with regular Cartesian sampling (31.56\% rate).
From left to right columns: GRAPPA(25.6656/0.4671/0.2494), SPIRiT(29.5550/0.6574/0.1594), VN (31.5546/0.7387/0.1333), deepcomplexMRI (38.6842/0.9360/0.0587),  pMRI-Net (38.8749/0.9375/0.0574), and ground truth (PSNR/SSIM/RMSE). From top to bottom rows: image, zoom-in views, and pointwise absolute error to ground truth.}
\label{PDFS}
\end{figure*}

\begin{figure*}
\centering
\includegraphics[width=0.16\linewidth, angle=180]{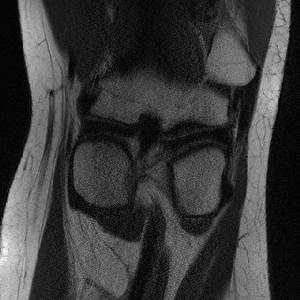}
\includegraphics[width=0.16\linewidth, angle=180]{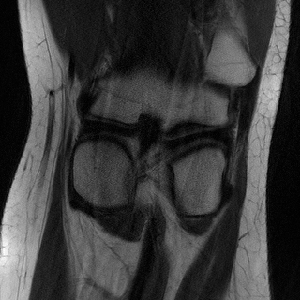}
\includegraphics[width=0.16\linewidth, angle=180]{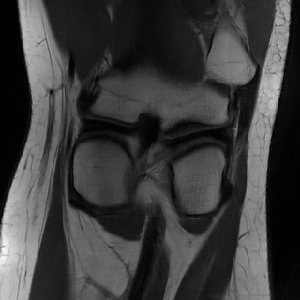}
\includegraphics[width=0.16\linewidth, angle=180]{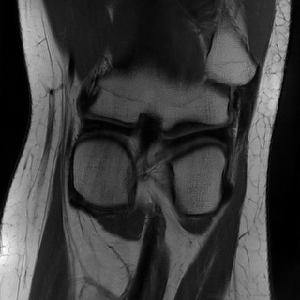}
\includegraphics[width=0.16\linewidth, angle=180]{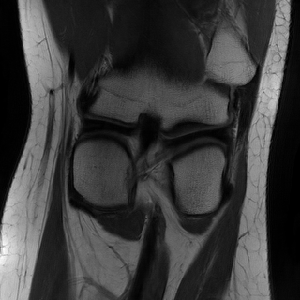}
\includegraphics[width=0.16\linewidth, angle=180]{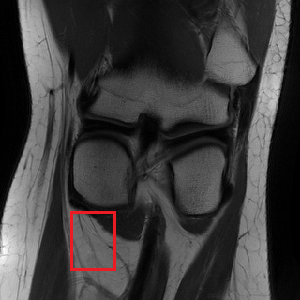}\\
\includegraphics[width=0.16\linewidth, angle=180]{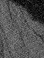}
\includegraphics[width=0.16\linewidth, angle=180]{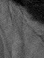}
\includegraphics[width=0.16\linewidth, angle=180]{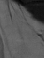}
\includegraphics[width=0.16\linewidth, angle=180]{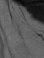}
\includegraphics[width=0.16\linewidth, angle=180]{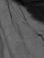}
\includegraphics[width=0.16\linewidth, angle=180]{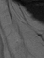}\\
\includegraphics[width=0.16\linewidth, angle=180]{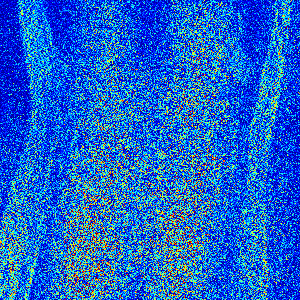}
\includegraphics[width=0.16\linewidth, angle=180]{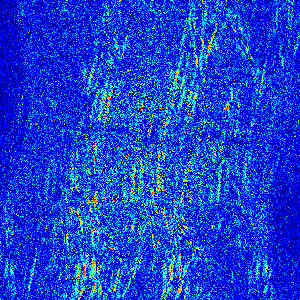}
\includegraphics[width=0.16\linewidth, angle=180]{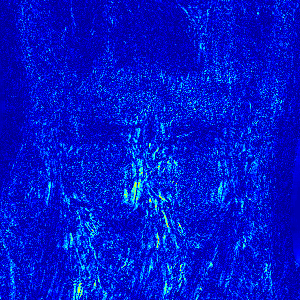}
\includegraphics[width=0.16\linewidth, angle=180]{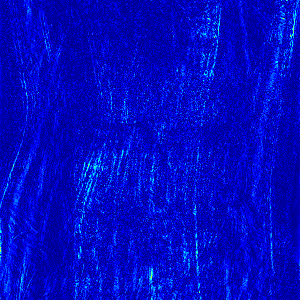}
\includegraphics[width=0.16\linewidth, angle=180]{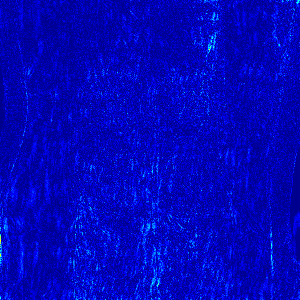}
\includegraphics[width=0.16\linewidth, angle=180]{fig/colorbar.png}
\caption{Results on the Coronal PD knee image with regular Cartesian sampling (31.56\% rate). From left to right columns: GRAPPA(29.9155/0.7360/0.1032), SPIRiT(33.2350/0.8461/0.0704), VN (38.3192/0.9464/0.0393), DeepcomplexMRI (41.2098/0.9713/0.0281), pMRI-Net (42.9330/0.9798/0.0231), and ground truth (PSNR/SSIM/RMSE). From top to bottom rows: image, zoom-in views, and pointwise absolute error to ground truth.}
\label{PD}
\end{figure*}
\textbf{Data.} Two sequences of data named Coronal proton-density (PD) and Coronal fat-saturated proton-density (FSPD) along with the regular Cartesian sampling mask with 31.56\% sampling ratio were obtained from \url{https://github.com/VLOGroup/mri-variationalnetwork} in our experiment. Each of the two sequences data were scanned from 20 patients. The training data consists of  526 central image slices with matrix size $ 320 \times 320$ from 19 patients, and we randomly pick 15 central image slices from the one patient that not included in training data as the testing data. We normalized training data by the maximum of the absolute valued zero-filled reconstruction.
\medskip

\noindent
\textbf{Implementation.} The proposed network was implemented with $T=5$ phases. We use Xavier initialization \cite{pmlr-v9-glorot10a} to initialize network parameters and Adam optimizer for training. Experiments apply mini-batches of 2 and 3000 epochs with learning rate 0.0001 and 0.0005 for training Coronal FSPD data and PD data respectively. The initial step size $ \rho_0 =0.1$, threshold parameter $\alpha_0 = 0$ and $\gamma=10^5$ in the loss function. All the experiments were implemented in TensorFlow on a workstation with Intel Core i9-7900 CPU and Nvidia GTX-1080Ti GPU.
\medskip

\noindent
\textbf{Evaluation.}
 We evaluate traditional methods GRAPPA \cite{griswold2002generalized}, SPIRiT \cite{doi:10.1002/mrm.22428}, and deep learning methods VN \cite{doi:10.1002/mrm.26977} , DeepcomplexMRI \cite{WANG2020136} over the 15 testing  Coronal  FSPD and  PD knee images in terms of PSNR, SSIM \cite{wang2004image} and RMSE (RMSE of $ \hat{\xbf}$ to true $\xbf^*$ is defined by $\| \hat{\xbf} - \xbf^*\|/\| \xbf^* \| $).
\medskip

\noindent
\textbf{Experimental results.} The average numerical performance with standard deviations are summarized in Table  \ref{tab:fs} and \ref{tab:pd}. The comparison on reconstructed images are shown in Fig.\ref{PDFS} and Fig.\ref{PD} for Coronal FSPD and PD testing data respectively.
Despite of the lack of coil sensitivities in training and testing, the proposed method still outperforms VN in reconstruction accuracy significantly while VN uses precomputed coil sensitivity maps from ESPIRiT\cite{uecker2014espirit}, which further shows that the proposed method can achieve improved accuracy without knowledge of coil sensitivities. Comparing 10 complex CNN blocks in
DeepcomplexMRI  with 5 phases in pMRI-Net, the latter requires fewer network parameters and less training time but improves reconstruction quality.

In the experiment of GRAPPA and SPIRiT, we use calibration kernel size $ 5\times 5$ with Tikhonov regularization in the calibration setted as 0.01. We implement SPIRiT with 30 iterations and set Tikhonov regularization in the reconstruction as $10^{-3}$. Default parameter settings for experiments of VN and DeepcomplexMRI were applied.  The final recovered image from VN is a full FOV single channel image, and DeepcomplexMRI produces a multi-coil image, which are combined into single channel image using adaptive multi-coil combination method \cite{Walsh2000682}. pMRI-Net reconstructs both single channel image $ \J( \ubf^{(T)}(\fbf; \Theta))$ and multi-channel image $\{\ubf_{i}^{(T)}(\fbf_{i}; \Theta)\} $.

\section{Conclusion}
\label{sec:conclusion}
We exploit a learning based multi-coil MRI reconstruction without explicit knowledge of coil sensitivity maps and the network is modeled in CS framework with proximal gradient scheme. The proposed network is designed to combine features of channel-wise images, and then extract sparse features from the coil combined image. Our experiments showed better performance of proposed ``combine-then-regularize” approach. 

%

\bibliographystyle{splncs04}
\bibliography{reference}

\title{Supplementary Material}

\author{
Wanyu Bian\inst{1} \and
Yunmei Chen\inst{1} \and
Xiaojing Ye\inst{2}}


\institute{University of Florida, Gainesville FL 32611, USA\\
\email{ \{wanyu.bian, yun\}@ufl.edu} \and
Georgia State University, Atlanta GA 30302, USA\\
\email{xye@gsu.edu}}
\maketitle

 \begin{figure}
\includegraphics[width=1\textwidth]{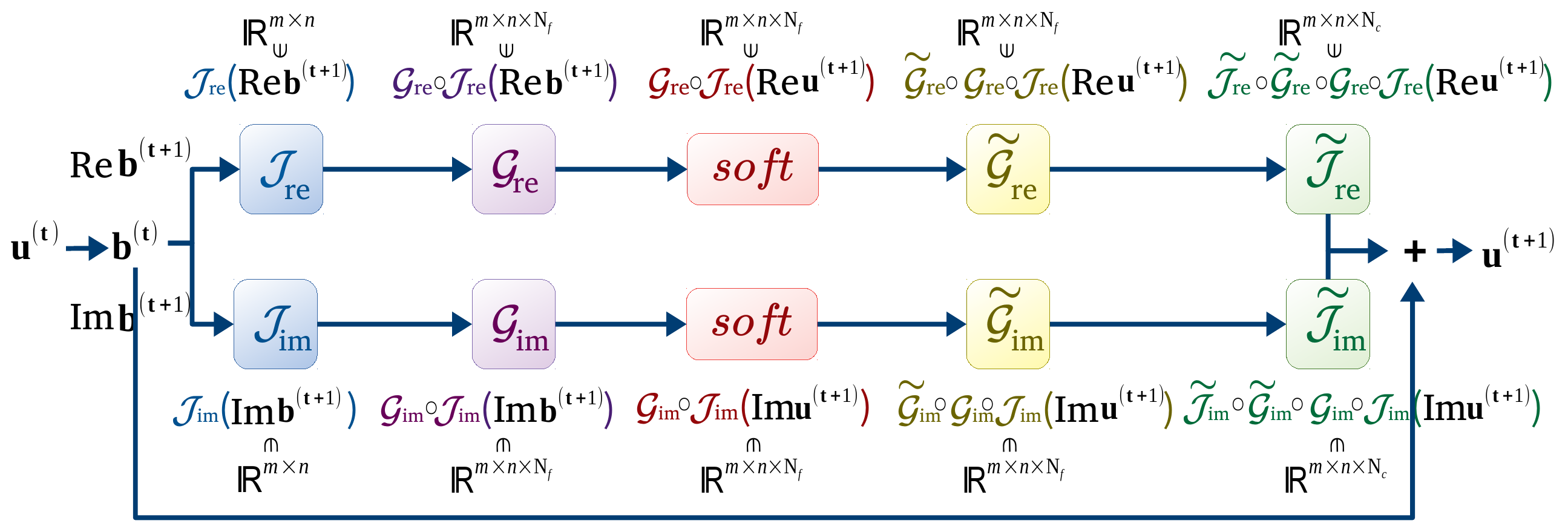}
\caption{Network structure paradigm in $t+1$th phase. $ \text{Re} \bbf^{(t+1)} , \text{Re} \ubf^{(t+1)} \in  \R^{ m \times n \times N_c}$ and $ \text{Im} \bbf^{(t+1)},  \text{Im} \ubf^{(t+1)} \in \R^{ m \times n \times N_c} $ represent for real and imaginary part of $ \bbf^{(t+1)} $ and $  \ubf^{(t+1)}$ respectively. The output of each real and imaginary convolutional operators corresponds to different colors. }
\label{fig9}
\end{figure}

\begin{figure}
\includegraphics[width=\textwidth]{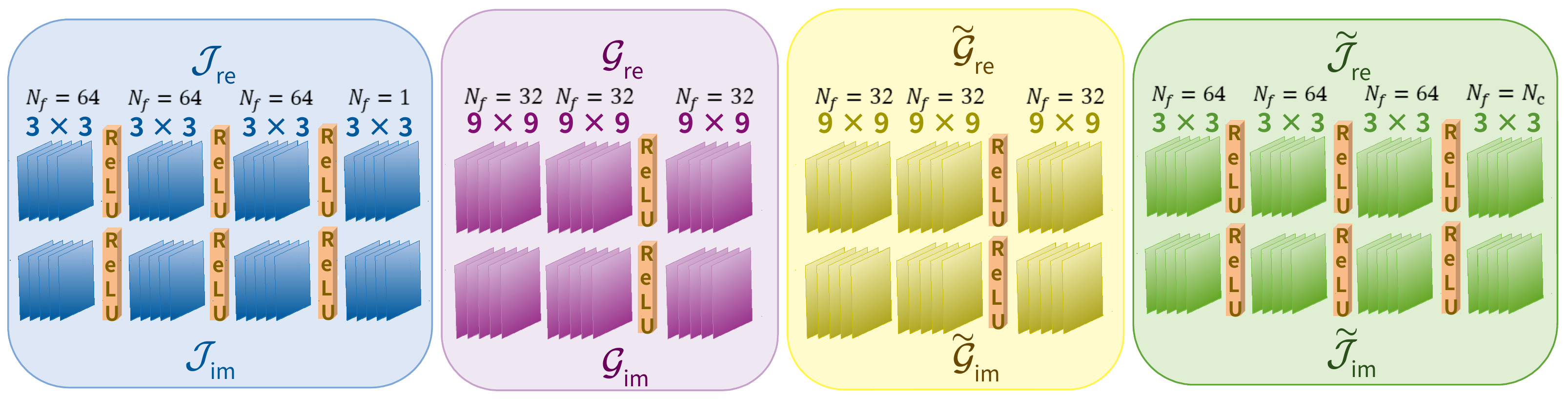}
\caption{ Structure of each convolutional operator. The weights of real and imaginary parts are not shared in the network, while both real and imaginary convolutional operators have the same structure.} \label{fig8}
\end{figure}

\begin{figure}[t]
\centering
\includegraphics[width=0.16\linewidth, angle=180]{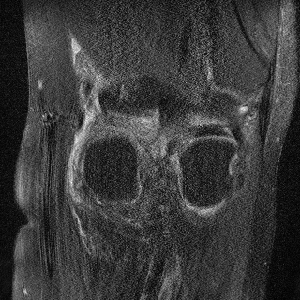}
\includegraphics[width=0.16\linewidth, angle=180]{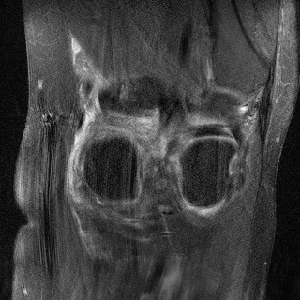}
\includegraphics[width=0.16\linewidth, angle=180]{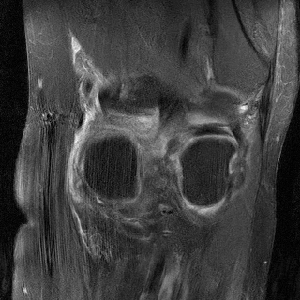}
\includegraphics[width=0.16\linewidth, angle=180]{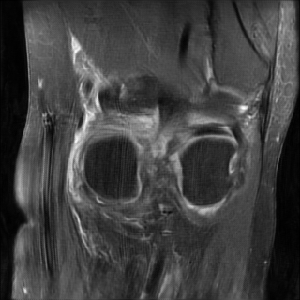}
\includegraphics[width=0.16\linewidth, angle=180]{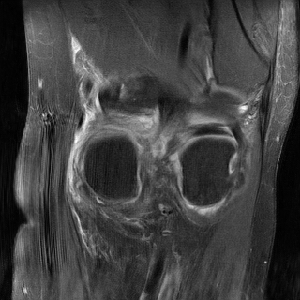}
\includegraphics[width=0.16\linewidth, angle=180]{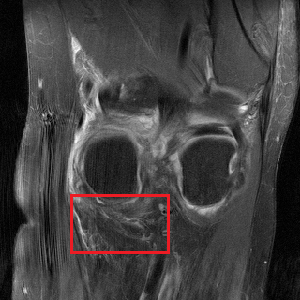}\\
\includegraphics[width=0.16\linewidth, angle=180]{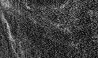}
\includegraphics[width=0.16\linewidth, angle=180]{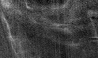}
\includegraphics[width=0.16\linewidth, angle=180]{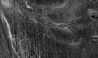}
\includegraphics[width=0.16\linewidth, angle=180]{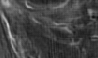}
\includegraphics[width=0.16\linewidth, angle=180]{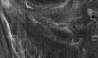}
\includegraphics[width=0.16\linewidth, angle=180]{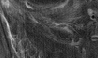}\\
\includegraphics[width=0.16\linewidth, angle=180]{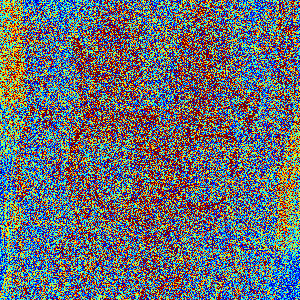}
\includegraphics[width=0.16\linewidth, angle=180]{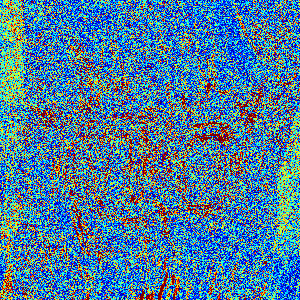}
\includegraphics[width=0.16\linewidth, angle=180]{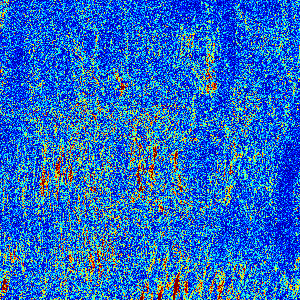}
\includegraphics[width=0.16\linewidth, angle=180]{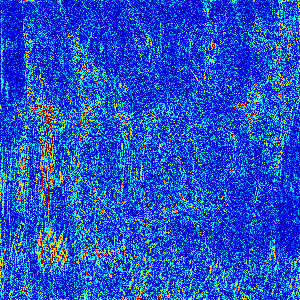}
\includegraphics[width=0.16\linewidth, angle=180]{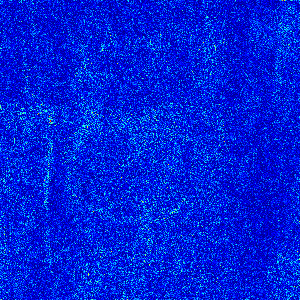}
\includegraphics[width=0.16\linewidth, angle=180]{fig/colorbar.png}
\caption{ Additional experiments results on the Coronal FSPD knee image with regular Cartesian sampling (31.56\% rate).
From left to right columns: GRAPPA(22.2203/0.3596/0.2666), SPIRiT(25.3434/0.5269/0.1861), VN(29.2276/0.7591/0.1190), DeepcomplexMRI(31.3597/0.8231/0.0931), proposed (36.8831/0.9372/0.0493), and ground truth (PSNR/SSIM/RMSE). From top to bottom rows: image, zoom-in views, and pointwise absolute error to ground truth.}
\label{fig7}
\end{figure}

\begin{figure}[h]
\centering
\includegraphics[width=0.16\linewidth, angle=180]{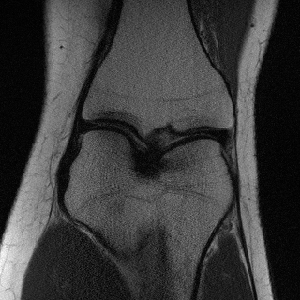}
\includegraphics[width=0.16\linewidth, angle=180]{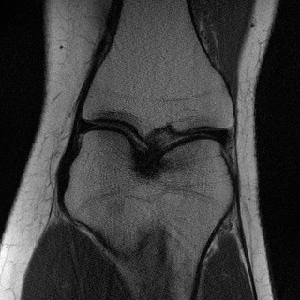}
\includegraphics[width=0.16\linewidth, angle=180]{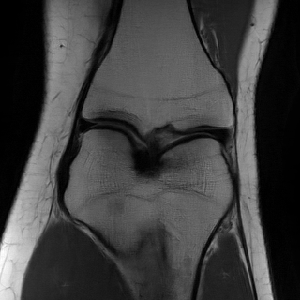}
\includegraphics[width=0.16\linewidth, angle=180]{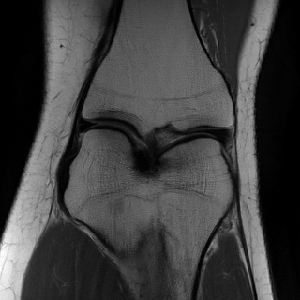}
\includegraphics[width=0.16\linewidth, angle=180]{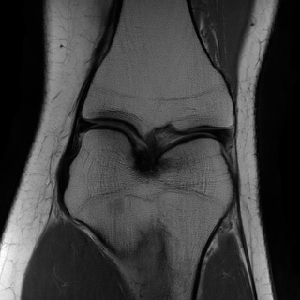}
\includegraphics[width=0.16\linewidth, angle=180]{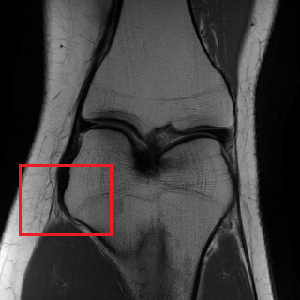}\\
\includegraphics[width=0.16\linewidth, angle=180]{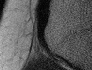}
\includegraphics[width=0.16\linewidth, angle=180]{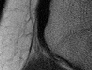}
\includegraphics[width=0.16\linewidth, angle=180]{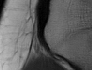}
\includegraphics[width=0.16\linewidth, angle=180]{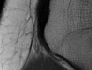}
\includegraphics[width=0.16\linewidth, angle=180]{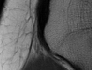}
\includegraphics[width=0.16\linewidth, angle=180]{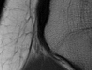}\\
\includegraphics[width=0.16\linewidth, angle=180]{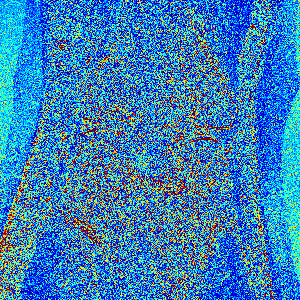}
\includegraphics[width=0.16\linewidth, angle=180]{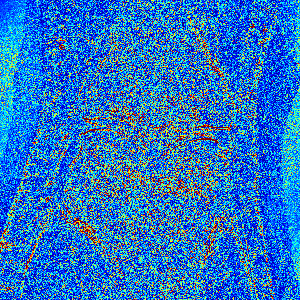}
\includegraphics[width=0.16\linewidth, angle=180]{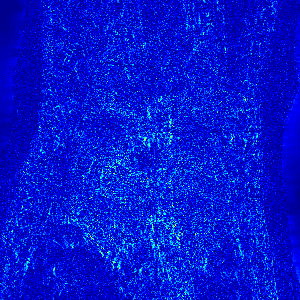}
\includegraphics[width=0.16\linewidth, angle=180]{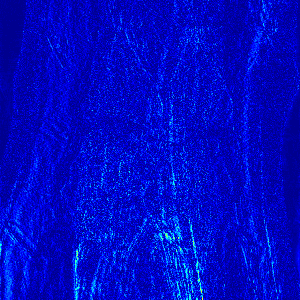}
\includegraphics[width=0.16\linewidth, angle=180]{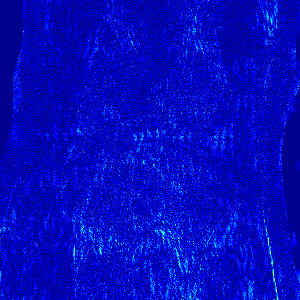}
\includegraphics[width=0.16\linewidth, angle=180]{fig/colorbar.png}
\caption{Additional experiments results on the Coronal PD knee image with regular Cartesian sampling (31.56\% rate). From left to right columns: GRAPPA(27.3571/0.5338/0.1357), SPIRiT(28.3076/0.5592/0.1216), VN (38.1679/0.9259/0.0391), DeepcomplexMRI (40.9388/0.9678/0.0284), Proposed (41.7025/0.9697/0.0260) and ground truth (PSNR/SSIM/RMSE). From top to bottom rows: image, zoom-in views, and pointwise absolute error to ground truth.}
\label{fig6}
\end{figure}

\end{document}